\documentstyle[aps,twocolumn]{revtex}

\begin{document}
\draft

\twocolumn[\hsize\textwidth\columnwidth\hsize\csname@twocolumnfalse\endcsname
\title{Continuous-variable and hybrid quantum gates}
\author{Xiaoguang Wang}
\address{Institute of Physics and Astronomy, Aarhus University, 
         DK-8000, Aarhus C, Denmark, and}
\address{Quantum information processing group, 
Institute for Scientific Interchange (ISI) Foundation, Viale Settimio Severo 65, 
I-10133 Torino, Italy}
\date{\today}
\maketitle

\begin{abstract}
We provide several schemes to construct the continuous-variable SWAP gate
and present a Hermitian generalized many-body continuous controlled$^{n}$-NOT gate. 
We introduce and study the hybrid controlled-NOT gate and controlled-SWAP gate,  
and physical realizations of them are discussed in trapped-ion systems. These 
continuous-variable and hybrid quantum gates may be used in the corresponding
continuous-variable and hybrid quantum computations.
\end{abstract}

]

\section{Introduction}

The quantum computer\cite{Feynman,Deutsch} is a device which operates with
quantum logic gates. It was shown that any quantum computation can be built
from a series of one-bit and two-bit quantum logic gates\cite{Gates}. The
fundamental controlled-NOT (CN)\cite{CN} gate, widely discussed in the
literature\cite{CNCN}, is the two-qubit gate in which one qubit is flipped
conditioned on the state of another qubit. Mathematically the CN gate is
defined as

\begin{equation}
\text{CN}_{12}|i\rangle _1|j\rangle _2=|i\rangle _1|i\oplus j\rangle _2,
\end{equation}
where $|i\rangle _1|j\rangle _2(i,j=0,1)$ are the basis states of the two
qubits, $\oplus $ denotes addition modulo 2. The first (second) qubit is the
control (target). 

It is known that an unknown qubit state $|\psi \rangle $ can be swapped with
the qubit state $|0\rangle $ using only two CN gates\cite{Xinlan}, i.e.,

\begin{equation}
\text{CN}_{21}\text{CN}_{12}|\psi \rangle _1|0\rangle _2=|0\rangle _1|\psi
\rangle _2.
\end{equation}
In Ref.\cite{Collins}, the gate CN$_{21}$CN$_{12}$ is called double CN gate.
Using the CN gates one can construct a general two-qubit SWAP gate as follows

\begin{equation}
\text{SWAP}_{12}=\text{CN}_{12}\text{CN}_{21}\text{CN}_{12},  \label{eq:cn3}
\end{equation}
which makes the transformation

\begin{equation}
\text{SWAP}_{12}|i\rangle _1|j\rangle _2=|j\rangle _1|i\rangle _2.
\end{equation}
The SWAP gate can be constructed in an alternative way as\cite{Heis}, 
\begin{equation}
\text{SWAP}_{12}\text{=}\frac 12\left( 1+\sigma _{x1}\sigma _{x2}+\sigma
_{y1}\sigma _{y2}+\sigma _{z1}\sigma _{z2}\right) ,  \label{eq:swap}
\end{equation}
where the operators $\sigma _{\alpha i}(\alpha =x,y,z)$ are the usual Pauli
operators of system $i.$ The remarkable properties of the SWAP gate are
described by Collins {\it et al.}\cite{Collins}, Eisert {\it et al.}\cite
{Eisert}, and Chefles {\it et al.}\cite{Chefles}. Both the CN gate and SWAP
gate are two-qubit gates. The one-qubit gates include NOT gate which is
expressed by the Pauli operator $\sigma _x$ and the Hadmard gate

\begin{equation}
\text{H}=\frac 1{\sqrt{2}}(\sigma _x+\sigma _z)
\end{equation}
which makes the transformation

\begin{mathletters}
\begin{eqnarray}
\text{H}|0\rangle &=&\frac 1{\sqrt{2}}(|0\rangle +|1\rangle ), \\
\text{H}|1\rangle &=&\frac 1{\sqrt{2}}(|0\rangle -|1\rangle ).  \label{eq:pm}
\end{eqnarray}
Both the NOT gate and the Hadmard gate are self-inverse, i.e., the square of
them are the identity operators.

For three qubits there are two types of gates, the Toffoli gate\cite{Toffoli}
and Fredkin gate\cite{Fredkin}, which are also called (controlled)$^2$-NOT
gate and the controlled-SWAP (CSWAP) gate, respectively. The CSWAP gate
makes the following transformation

\end{mathletters}
\begin{mathletters}
\begin{eqnarray}
\text{CSWAP}_{(12)3}|i\rangle _1|j\rangle _2|0\rangle _3 &=&|i\rangle
_1|j\rangle _2|0\rangle _3, \\
\text{CSWAP}_{(12)3}|i\rangle _1|j\rangle _2|1\rangle _3 &=&|j\rangle
_1|i\rangle _2|1\rangle _3,
\end{eqnarray}
where the third qubit acts as the control. The quantum gates described above
act on discrete variables, the qubits. In this paper we give the
continuous-variable and hybrid versions of quantum gates, which may be used
in the continuous-variable\cite{Lloyd} and hybrid\cite{Hybrid} quantum
computation. In the hybrid version of quantum gates the discrete variable
acts as the control and the continuous variables as the targets.

In Sec. II we begin with the introduction of the one-body gates for
continuous variables. We proceed in Sec. III to study the two-body and
many-body continuous-variable gates and consider the CN gate, SWAP\ gate,
and controlled$^n$-NOT gate as well as the cloning gate. Several methods are
proposed to realize the SWAP gate. In Sec. III we introduce and study the
hybrid quantum gates, hybrid CN gates and CSWAP gates. We give two schemes
to realize the hybrid gates in trapped-ion systems. The conclusion is given
in Sec. V.

\section{One-body gates for continuous variables}

\subsection{NOT gate}

The one-body continuous-variable NOT gate may be defined as  the parity
operator$\,$

\end{mathletters}
\begin{equation}
\text{NOT}=(-1)^{a^{\dagger }a},\,
\end{equation}
where $a$ and $a^{\dagger }$ are bosonic annihilation and creation
operators. It is easy to see that

\begin{eqnarray}
\text{NOT}|x\rangle &=&|-x\rangle ,  \nonumber \\
\text{NOT}|p\rangle &=&|-p\rangle ,  \nonumber \\
\text{NOT}^2 &=&1,
\end{eqnarray}
where $|x\rangle $ is the eigenstate of the position operator $\hat{x},$ and 
$|p\rangle $ is the eigenstate of the momentum operator $\hat{p}.$

\subsection{Hadamard gate}

The continuous version of the Hadmard gate is in fact the Fourier
transformation and defined by\cite{CNplus}

\begin{equation}
\text{F(}\sigma \text{)}|x\rangle =\frac 1{\sigma \sqrt{\pi }}\int
dye^{2ixy/\sigma ^2}|y\rangle ,
\end{equation}
where $\sigma $ is the scaled length. This is the transformation used to go
from the position to the momentum basis if we set $\sigma =\sqrt{2}.$ The
inverse F$^{\dagger }$($\sigma $) is obtained by replacing $i$ by $-i$
giving the result that

\begin{equation}
\text{F(}\sigma \text{)F}^{\dagger }(\sigma )|x\rangle =\text{F}^{\dagger
}(\sigma )\text{F(}\sigma \text{)}|x\rangle =|x\rangle .
\end{equation}
Note that the continuous-variable Hadamard gate is not self-inverse.

\section{Two-body and many body gates for continuous variables}

\subsection{CN gate}

The two-qubit CN gate has been extended to the case of continuous variables,
the gates CN$_{12}^{+}$\cite{CNplus} and CN$_{12}^{-}\cite{CNminus},$ which
are defined by

\begin{eqnarray}
\text{CN}_{12}^{\pm }|x\rangle _1|y\rangle _2 &=&|x\rangle _1|x\pm y\rangle
_2, \\
\text{CN}_{12}^{+} &=&e^{-i\hat{x}_1\hat{p}_2},  \label{eq:cnplus} \\
\text{CN}_{12}^{-} &=&\text{NOT}_2e^{i\hat{x}_1\hat{p}_2}=e^{-i\hat{x}_1\hat{%
p}_2}\text{NOT}_2,  \label{eq:cnminus}
\end{eqnarray}
where the position operator of system $i$ $(i=1,2)$ is denoted by $\hat{x}%
_i\,$and the momentum operator by $\hat{p}_i.$ In momentum space the CN gate
can be defined as

\begin{eqnarray}
\text{CN}_{12}^{\pm }|p\rangle _1|q\rangle _2 &=&|p\rangle _1|p\pm q\rangle
_2, \\
\text{CN}_{12}^{+} &=&e^{i\hat{x}_2\hat{p}_1}, \\
\text{CN}_{12}^{-} &=&\text{NOT}_2e^{-i\hat{x}_2\hat{p}_1}=e^{i\hat{x}_2\hat{%
p}_1}\text{NOT}_2.  \label{eq:cnmp}
\end{eqnarray}
The definitions of the CN gates are basis dependent. From Eqs.(\ref
{eq:cnplus}), (\ref{eq:cnminus}), and (\ref{eq:cnmp}), it is easy to check
that both gates are unitary, the gate CN$_{12}^{+}$ is not Hermitian and not
self-inverse, while CN$_{12}^{-}$ is Hermitian and self-inverse.

The CN gate for qubits has been used in various kinds of quantum information
processing such as teleportation\cite{Tele}, dense coding\cite{Dense},
quantum state swapping\cite{CN}, entangling quantum states\cite{Entangle}
and Bell measurements\cite{Bell}. It is natural to ask that if the
continuous CN gates can perform some similar tasks like entangling and
swapping quantum states. Let the continuous CN gates CN$_{12}^{\pm }$ and
the Hadamard gate F($\sqrt{2}$) act on the state $|z\rangle _1|y\rangle _2.$
The resultant states are entangled states

\begin{eqnarray}
|\psi \rangle ^{\pm } &=&\text{CN}_{12}^{\pm }\text{F(}\sqrt{2}\text{)}%
|z\rangle _1|y\rangle _2  \nonumber \\
&=&\frac 1{\sqrt{2\pi }}\int dxe^{ixz}|x\rangle _1|x\pm y\rangle _2.
\end{eqnarray}
It is interesting to see that the following equations

\begin{mathletters}
\begin{eqnarray}
(\hat{x}_1-\hat{x}_2)|\psi \rangle ^{\pm } &=&\mp y|\psi \rangle ^{\pm }, \\
(\hat{p}_1+\hat{p}_2)|\psi \rangle ^{\pm } &=&z|\psi \rangle ^{\pm }
\end{eqnarray}
hold. That is to say, both the entangled states $|\psi \rangle ^{\pm }$ are
the common eigenvectors of the position difference operator $\hat{x}_1-\hat{x%
}_2$ and momentum sum operator $\hat{p}_1+\hat{p}_2.$ Further both the
continuous CN gates can be used to construct $N$-party entangled state as
follows

\end{mathletters}
\begin{eqnarray}
\text{CN}_{12}^{\pm }\text{CN}_{13}^{\pm }...\text{CN}_{1N}^{\pm }|p
&=&0\rangle _1|x=0\rangle _2|x=0\rangle _3...|x=0\rangle _N  \nonumber \\
&=&\frac 1{\sqrt{2\pi }}\int dx|x\rangle _1|x\rangle _2...|x\rangle _N.
\end{eqnarray}
This state is obtained by Braunstein\cite{CNplus} by a series of beam
splitters. Here we provide an alternative way to obtain this state by using $%
N$ CN gates. The $N$-party entangled state is an eigenstate with total
momentum zero and relative positions zero.

\subsection{SWAP gate}

Having seen that both the continuous CN gates can entangle quantum states,
then we ask if they can perform quantum state swapping by certain
combinations of them. For continuous variables we have

\begin{equation}
\text{CN}_{21}^{-}\text{CN}_{12}^{\pm }|x\rangle _1|y=0\rangle
_2=|y=0\rangle _1|x\rangle _2.
\end{equation}
From Eq.(\ref{eq:cn3}), one may guess that similar expression exists for
continuous-variable SWAP gate. It is straightforward to check that

\begin{mathletters}
\begin{eqnarray}
\text{CN}_{12}^{+}\text{CN}_{21}^{+}\text{CN}_{12}^{+}|x\rangle _1|y\rangle
_2 &=&|2x+y\rangle _1|3x+2y\rangle _2{\bf ,} \\
\text{CN}_{12}^{-}\text{CN}_{21}^{-}\text{CN}_{12}^{-}|x\rangle _1|y\rangle
_2 &=&|-y\rangle _1|-x\rangle _2{\bf .}
\end{eqnarray}
Then the SWAP gate can be constructed as

\end{mathletters}
\begin{eqnarray}
\text{SWAP}_{12} &=&\text{NOT}_1\text{NOT}_2\text{CN}_{12}^{-}\text{CN}%
_{21}^{-}\text{CN}_{12}^{-}  \nonumber \\
&=&\text{CN}_{12}^{-}\text{CN}_{21}^{-}\text{CN}_{12}^{-}\text{NOT}_1\text{%
NOT}_2, \\
\text{SWAP}_{12}|x\rangle _1|y\rangle _2 &=&|y\rangle _1|x\rangle _2.
\end{eqnarray}
We see that one can not obtain the SWAP{\bf \ }gate by only the gates CN$%
_{ij}^{+}(i\neq j),$ while one can use the gates CN$_{ij}^{-}$ to obtain it.
Different from the situation of discrete variables, here the
continuous-variable SWAP gate needs two NOT gates$.\,$In fact the gates CN$%
_{ij}^{+}(i\neq j)$ is not completely useless in the realization of the SWAP
gate. Using both the gates CN$_{ij}^{+}$ and CN$_{ij}^{-}$, we have

\begin{eqnarray}
\text{SWAP}_{12} &=&\text{NOT}_2\text{CN}_{12}^{-}\text{CN}_{21}^{-}\text{CN}%
_{12}^{+}  \nonumber \\
&=&e^{i\hat{x}_1\hat{p}_2}\text{NOT}_1e^{i\hat{x}_2\hat{p}_1}e^{-i\hat{x}_1%
\hat{p}_2}.  \label{eq:swapswap}
\end{eqnarray}
Here we have used Eqs.(\ref{eq:cnplus}) and (\ref{eq:cnminus}). Then we can
construct the SWAP gate using one-body gate and three two-body gates. The
SWAP gate acting on momentum space can be constructed similarly.

Recalling that the two-qubit SWAP gate can be given in Eq.(\ref{eq:swap}),
we expect that the continuous SWAP gate be implemented in another way. Now
we introduce the operator

\begin{equation}
B_{12}=e^{i\frac \pi 2(\hat{x}_1\hat{p}_2-\hat{x}_2\hat{p}_1)}
\label{eq:sw1}
\end{equation}
acting on the two continuous systems 1 and 2. The operator corresponds to a
beam splitter and makes the transformation 
\begin{equation}
B_{12}\left( 
\begin{array}{l}
\hat{p}_1 \\ 
\hat{p}_2
\end{array}
\right) B_{12}^{\dagger }=\left( 
\begin{array}{l}
-\hat{p}_2 \\ 
\hat{p}_1
\end{array}
\right) ,  \label{eq:swsw}
\end{equation}
from which we have

\begin{equation}
B_{12}|x\rangle _1|y\rangle _2=|y\rangle _1|-x\rangle _2.
\end{equation}
Then the continuous-variable SWAP gate is immediately obtained as

\begin{equation}
\text{SWAP}_{12}=\text{NOT}_2B_{12}.  \label{eq:sw2}
\end{equation}
From Eqs.(\ref{eq:swsw}) and (\ref{eq:sw2}), the swapping function of the
SWAP gate can be compactly stated by

\begin{eqnarray}
\text{SWAP}_{12}\left( 
\begin{array}{l}
\hat{p}_1 \\ 
\hat{p}_2
\end{array}
\right) \text{SWAP}_{12} &=&\left( 
\begin{array}{l}
\hat{p}_2 \\ 
\hat{p}_1
\end{array}
\right) ,  \nonumber \\
\text{SWAP}_{12}\left( 
\begin{array}{l}
\hat{x}_1 \\ 
\hat{x}_2
\end{array}
\right) \text{SWAP}_{12} &=&\left( 
\begin{array}{l}
\hat{x}_2 \\ 
\hat{x}_1
\end{array}
\right) ,
\end{eqnarray}
which may serve as alternative definitions.

Substituting $\hat{x}_j=\frac 1{\sqrt{2}}(a_j+a_j^{\dagger }),\hat{p}_j=%
\frac 1{i\sqrt{2}}(a_j-a_j^{\dagger })$ to the Eq.(\ref{eq:sw1}), we can
reexpress the operator $B_{12}$ in terms of the annihilation and creation
operators and then rewrite the SWAP gate (\ref{eq:sw2}) as

\begin{equation}
\text{SWAP}_{12}=e^{i\pi a_2^{\dagger }a_2}e^{\frac \pi 2(a_1^{\dagger
}a_2-a_2^{\dagger }a_1)}.
\end{equation}
Let the above SWAP gate act on the discrete Fock basis states, we obtain 
\begin{equation}
\text{SWAP}_{12}|n\rangle _1|m\rangle _2=|m\rangle _1|n\rangle _2,
\label{eq:Fock}
\end{equation}
where $|n\rangle _{i\text{ }}$denotes the Fock state of system $i.\,$Eq.(\ref
{eq:Fock}) in fact gives the representation of the SWAP gate in the two-mode
Fock space. We see that the SWAP gate is basis independent, while the CN
gate is basis dependent.

As an end of this subsection mention a relation between the SWAP gate and
the CN gates,

\begin{equation}
\text{SWAP}_{12}\text{CN}_{12}\text{SWAP}_{12}=\text{CN}_{21}.
\end{equation}
The above equation shows that one can use the SWAP gate and CN gate CN$_{12}$
to realize another CN gate CN$_{21}.$

\subsection{ Controlled$^n$-NOT gate}

We define a Hermitian continuous generalization of the discrete controlled$%
^n $- NOT gate as

\begin{eqnarray}
&&\text{CN}_{(12...N)N+1}^{\pm }|x_1\rangle _1|x_2\rangle _2...|x_N\rangle
_N|x_{N+1}\rangle _{N+1}  \nonumber \\
&=&|x_1\rangle _1|x_2\rangle _2...|x_N\rangle
_N|-x_{N+1}+\sum_{n=1}^Nx_n\rangle _{N+1},
\end{eqnarray}
\begin{equation}
\text{CN}_{(12...N)N+1}^{\pm }=\text{NOT}_{N+1}e^{i\hat{p}_{N+1}\sum_{n=1}^N%
\hat{x}_n},
\end{equation}
Similar gate can be defined in momentum space. Then the gate defined in this
way is both unitary and Hermitian, and therefore self-inverse. For the case $%
N=2$ and $3,$ the gate becomes the continuous-variable CN and Toffoli gate,
respectively.

\subsection{1$\rightarrow 2$ cloning gate}

For discrete variables the CN gates CN$_{21}$ and CN$_{31\text{ }}$commute
with each other, however for continuous variables, from Eq.(\ref{eq:cnminus}%
), the following equation

\begin{equation}
\lbrack \text{CN}_{31}^{-},\text{CN}_{21\text{ }}^{-}]=e^{i(\hat{x}_2-\hat{x}%
_3)\hat{p}_1}-e^{i(\hat{x}_3-\hat{x}_2)\hat{p}_1}  \label{eq:cncn}
\end{equation}
holds for two Hermitian CN gates CN$_{21}^{-}$ and CN$_{31\text{ }}^{-}$.
That is to say, these two continous--variable CN gates do not commute.

It is known that the 1$\rightarrow 2$ cloning gate is described by\cite
{Buzek2}

\begin{equation}
{\cal C}=\text{CN}_{31}\text{CN}_{21}\text{CN}_{13}\text{CN}_{12}
\end{equation}
in terms of four CN gates. To generalize directly to the continuous case of
the above cloning gate, we obtain

\begin{equation}
{\cal C}^{\prime }=\text{CN}_{31}^{-}\text{CN}_{21}^{-}\text{CN}_{13}^{-}%
\text{CN}_{12}^{-}.
\end{equation}
Using Eqs.(\ref{eq:cnminus}) and (\ref{eq:cncn}), we rewrite the gate ${\cal %
C}^{\prime }$ as

\begin{equation}
{\cal C}^{\prime }=e^{-i(\hat{x}_3-\hat{x}_2)\hat{p}_1}e^{-i\hat{x}_1(\hat{p}%
_2+\hat{p}_3)}\text{NOT}_2\text{NOT}_3,
\end{equation}
which is just the continuous-variable 1$\rightarrow 2$ cloning gate up to
the two NOT gates\cite{Cerf}.

\section{Hybrid gates}

Now we introduce and study two kinds of hybrid quantum gates, the hybrid CN
gate and CSWAP gate.

\subsection{Hybrid CN gate}

We define the hybrid CN gate as

\begin{eqnarray*}
\text{CN}_{12}^{\prime }|0\rangle _1|x\rangle _2 &=&|0\rangle _1|x\rangle _2,
\\
\text{CN}_{12}^{\prime }|1\rangle _1|x\rangle _2 &=&|1\rangle _1|-x\rangle
_2,
\end{eqnarray*}
which can be realized in a trapped-ion system. In trapped-ion systems, one
can have the following Hamiltonian experimentally\cite{Gerry97,Monroe}

\begin{equation}
H_1=\lambda a^{\dagger }a{\cal P}_1
\end{equation}
where $a$ and $a^{\dagger }$ are bosonic annihilation and creation operators
of the center-of-mass motion of the trapped ion, $\,{\cal P}_1=|1\rangle
_1\langle 1|$ is the projection operator, and $\lambda $ is the effective
coupling constant. It is easy to show that the evolution operator $%
e^{-i\lambda ta^{\dagger }a{\cal P}_1}$ at time $t=\pi /\lambda $ gives
directly the hybrid CN gate. One simple application of this gate is the
generation of even and odd coherent states. Let the input state be $\frac 1{%
\sqrt{2}}(|0\rangle _1+|1\rangle _1)|\alpha \rangle _2$, where $|\alpha
\rangle _2$ is a bosonic coherent state. Then after the gate operation the
output state will be $\frac 1{\sqrt{2}}(|0\rangle _1|\alpha \rangle
_2+|1\rangle _1|-\alpha \rangle _2).$ Now we measure the qubit on the state $%
|\pm \rangle =\frac 1{\sqrt{2}}(|0\rangle \pm |1\rangle ),$ the continuous
state will collapse into the even and odd coherent states, respectively.

\subsection{Hybrid controlled-SWAP gate}

A general controlled-SWAP gate is described by the following transformation

\begin{eqnarray}
|\Psi \rangle _1|\Phi \rangle _2|0\rangle _3 &\rightarrow &|\Psi \rangle
_2|\Phi \rangle _1|0\rangle _3,  \nonumber \\
|\Psi \rangle _1|\Phi \rangle _2|1\rangle _3 &\rightarrow &|\Phi \rangle
_2|\Psi \rangle _1|1\rangle _3,  \label{eq:cswap}
\end{eqnarray}
This gate has three inputs and the third is the control qubit. Let the input
state of the CSWAP gate is $\frac 1{\sqrt{2}}|\Psi \rangle _1|\Phi \rangle
_2(|0\rangle _3+|1\rangle _3)$ and measure the output state. If we measure
the qubit on the state $|\pm \rangle _3=\frac 1{\sqrt{2}}(|0\rangle _3\pm
|1\rangle _3),$ we obtain exactly the symmetric and antisymmetric entangled
states, $|\Psi \rangle _1|\Phi \rangle _2\pm |\Phi \rangle _2|\Psi \rangle
_1 $ up to normalization constants. This is actually a universal entangler%
\cite{Buzek1}. So it is desirable to consider the CSWAP gate of the form (%
\ref{eq:cswap}) when then states $|\Psi \rangle _1$ and $|\Phi \rangle _2$
are continuous-variable states.

From the continuous-variable SWAP gate (\ref{eq:swapswap}), the CSWAP gate
is formally constructed as

\begin{equation}
\text{CSWAP}_{12(3)}^{\prime }=e^{i\hat{x}_1\hat{p}_2{\cal P}_3}e^{i\pi
a_1^{\dagger }a_1{\cal P}_3}e^{i\hat{x}_2\hat{p}_1{\cal P}_3}e^{-i\hat{x}_1%
\hat{p}_2{\cal P}_3}.  \label{eq:sw}
\end{equation}
where$\,{\cal P}_3=|1\rangle _3\langle 1|$ is the projection operator of the
control system 3. There are three three-body interactions in the expression
of the CSWAP gate. We will realize the CSWAP gate by two-body interactions.

First we see that the operators $e^{\pm ix\hat{p}}$ and $e^{\pm ip\hat{x}}$
satisfy relation

\begin{equation}
e^{ixp}=e^{ix\hat{p}}e^{ip\hat{x}}e^{-ix\hat{p}}e^{-ip\hat{x}}.
\end{equation}
The above relation can be generalized as \cite{Wang1}

\begin{eqnarray}
e^{ixp\sin \theta } &=&e^{i(\frac \pi 2-\theta )a^{\dagger }a}e^{ix\hat{p}%
}e^{-i(\frac \pi 2-\theta )a^{\dagger }a}e^{ip\hat{x}}  \nonumber \\
&&\times e^{i(\frac \pi 2-\theta )a^{\dagger }a}e^{-ix\hat{p}}e^{-i(\frac \pi
2-\theta )a^{\dagger }a}e^{-ip\hat{x}}.  \label{eq:ggg}
\end{eqnarray}
As the operator $\hat{p}_1,\,\hat{x}_2$ and ${\cal P}_{3\text{ }}$commutes
with each other, we replace $x$ with $\hat{x}_2$ , $p$ with $\hat{p}_1,$ and 
$\theta $ with $\pi {\cal P}_{3\text{ }}/2$ in Eq.(\ref{eq:ggg}),
respectively. Then we obtain

\begin{eqnarray}
e^{i\hat{p}_1\hat{x}_2{\cal P}_3} &=&e^{i\frac \pi 2(1-{\cal P}_3)a^{\dagger
}a}e^{i\hat{x}_2\hat{p}}e^{-i\frac \pi 2(1-{\cal P}_3)a^{\dagger }a}e^{i\hat{%
p}_1\hat{x}}  \nonumber \\
&&\times e^{i\frac \pi 2(1-{\cal P}_3)a^{\dagger }a}e^{-i\hat{x}_2\hat{p}%
}e^{-i\frac \pi 2(1-{\cal P}_3)a^{\dagger }a}e^{-i\hat{p}_1\hat{x}}
\end{eqnarray}
The above equation shows that we have written the three-body unitary
operator $e^{i\hat{p}_1\hat{x}_2{\cal P}_3}$ in terms of eight two-body
operators. Therefore the CSWAP gate (\ref{eq:sw}) can be written in terms of
two-body operators.

From Eqs. (\ref{eq:sw1}) and (\ref{eq:sw2}), we write the CSWAP gate as the
form

\begin{equation}
\text{CSWAP}_{12}^{\prime }=e^{i\pi a_2^{\dagger }a_2{\cal P}_3}e^{i\frac \pi
2(\hat{x}_1\hat{p}_2-\hat{x}_2\hat{p}_1){\cal P}_3},  \label{eq:sss}
\end{equation}
which also includes a three-body operator. Next we see how to realize this
CSWAP gate in a trapped-ion system.

Gerry derived an effective Hamiltonian for two modes $a$ and $b$ as\cite
{Gerry1}

\begin{equation}
H_2=\chi (a_1^{\dagger }a_1-a_2^{\dagger }a_2){\cal P}_3
\end{equation}
in a trapped-ion system. The Hamiltonian $H_2$ can be rewritten as

\begin{equation}
H_2=2\chi J_z{\cal P}_3,
\end{equation}
where $J_z=\frac 12(a_1^{\dagger }a_1-a_2^{\dagger }a_2).\,$The operators $%
J_z$, $J_{+}=a_1^{\dagger }a_2,$ and $J_{-}=a_2^{\dagger }a_1$ form the
su(2) Lie algebra. The unitary operator at time $t=-\pi /(2\chi )$
corresponds to the Hamiltonian is given by

\begin{equation}
U=e^{i\pi J_z{\cal P}_3}
\end{equation}
The unitary operator $U$ can be transformed to $U^{\prime }\,$as

\begin{eqnarray}
&&U^{\prime }=e^{i\frac \pi 2J_x}Ue^{-i\frac \pi 2J_x}  \nonumber \\
&=&e^{i\pi J_y{\cal P}_3}=e^{i\frac \pi 2(\hat{x}_1\hat{p}_2-\hat{x}_2\hat{p}%
_1){\cal P}_3},  \label{eq:ss}
\end{eqnarray}
where $J_x=(J_{+}+J_{-})/2$ and $J_y=\left( J_{+}-J_{-}\right) /(2i).$ From
Eqs. (\ref{eq:sss}) and (\ref{eq:ss}), we write the CSWAP gate as

\begin{eqnarray}
\text{C-SWAP}_{12}^{\prime } &=&e^{i\pi a_2^{\dagger }a_2{\cal P}_3}e^{i%
\frac \pi 4(a_1^{\dagger }a_2+a_2^{\dagger }a_1)}  \nonumber \\
&&\times e^{i\frac \pi 2a_1^{\dagger }a_1{\cal P}_3}e^{-i\frac \pi 2%
a_2^{\dagger }a_2{\cal P}_3}e^{-i\frac \pi 4(a_1^{\dagger }a_2+a_2^{\dagger
}a_1)}  \nonumber \\
&=&e^{i\frac \pi 2a_2^{\dagger }a_2{\cal P}_3}e^{-i\frac \pi 2a_1^{\dagger
}a_1{\cal P}_3}e^{i\frac \pi 4(a_1^{\dagger }a_2+a_2^{\dagger }a_1)} 
\nonumber \\
&&\times e^{i\pi a_1^{\dagger }a_1{\cal P}_3}e^{-i\frac \pi 4(a_1^{\dagger
}a_2+a_2^{\dagger }a_1)}.
\end{eqnarray}
Therefore we have given the form of CSWAP gate in terms of five two-body
operators.

We have used two methods to express the three-body hybrid CSWAP gate in
terms of two-body operators. In other words we provide two ways to realize
the CSWAP gate.

\section{Conclusion}

In conclusion we have introduced and studied the continuous and hybrid
versions of quantum gates. The continuous-variable gates include one-body
(NOT, Hadmard), two-body (CN, double CN, SWAP) and many-body gates
(controlled$^n$-NOT). Some relations between the CN, double CN and the SWAP
gates are given. The hybrid quantum gates include the hybrid CN gate and the
three-body controlled-SWAP gate. We proposed physical schemes to realize the
hybrid gates in the trapped-ion systems. It is interesting to see that most
of the quantum gates are not only unitary, but also Hermitian, and therefore
self-inverse.

\acknowledgments

The author thanks for the many helpful discussions with Klaus M\o lmer and
Anders S\o rensen. This work is supported by the Information Society
Technologies Programme IST-1999-11053, EQUIP, action line 6-2-1 and European
project Q--ACTA.

\end{document}